\documentclass{raa}            

\usepackage{graphicx,times}             
\input{epsf.sty}                        

\begin{document}
   \title{On the Lorentz Factor of Superluminal Sources}

   \volnopage{Vol.0 (200x) No.0, 000--000}      
   \setcounter{page}{1}          

   \author{C.C Onuchukwu
      \inst{1,2}
   \and A.A Ubachukwu
      \inst{2}
   }

   \institute{Department of Industrial Physics, Anambra State UNiversity Uli, Nigeria; {\it onuchukwu71chika@yahoo.com}\\
        \and
             Department of Physics and Astronomy, University of Nigeria Nsukka, Nigeria\\
   }

   \date{Received~~2012 month day; accepted~~2012~~month day}

\abstract {We investigate the properties of features seen within superluminal sources often referred to as components. Our result indicates a fairly strong correlation of $ r{\sim} 0.6$ for quasars, $ r{\sim} 0.4$ for galaxies, and $ r {\sim} 0.8$ for BL Lac objects in our sample between component sizes and distances from the stationary core. Assumption of free adiabatic expanding plasma enabled us to constrain in general the Lorentz factor for superluminal sources. Our estimated Lorentz factor of ${\gamma} {\sim} 7 -17$ for quasars,  ${\gamma}{\sim} 6 -13$ for galaxies and ${\gamma}{\sim} 4 - 9$ for BL Lac objects indicate that BL Lac have the lowest range of Lorentz factor.
}

  \authorrunning{C.C. Onuchukwu \& A.A. Ubachukwu }            
   \titlerunning{Lorentz Factor of Superluminal Sourecs }  

   \maketitle

%
%
\section{Introduction}           
\label{sect:intro}
A large number of superluminal motions have been observed in jets of numerous classes of astrophysical objects using Very Long Baseline Interferometer (VLBI). The most prominent sources of these jets are the Active Galactic Nuclei (AGN), where the jet emission contributes significantly to the spectrum of the source (K\"{o}rding \& Falcke 2004). These extragalactic radio jets have been intensively studied for a number of important reasons. These include their association with super massive accretion   black hole, which plays a substantial role in the formation and evolution of galaxies $-$ the building block of the universe.  Studies and applications of the astrophysics of these radio jets can help reveal black holes energetic and their interactions with their environments. Understanding the formation, collimation, acceleration, and propagation of these radio jets will aid our understanding of the laws of physics under extreme conditions $-$ relativistic speed, high magnetic field strength, etc. 

In the recent years, VLBI observations of AGN jets (e.g. see Kellermann et al. 2004; Britzen et al. 2007, 2008; the MOJAVE program; the TANAMI project) have provided us with observations for morphological studies and motions of features (often refer to as components) seen within these jets. These observations provide us with many values of the apparent luminosity, size, radial distance away from the core and proper motion/apparent transverse speed of components moving relativistically along the jets. Though these quantities are of considerable interest, the intrinsic properties (Lorentz factor $-$ bulk speed/pattern speed, angle to the line of sight) are more fundamental and necessary for kinematic characterization of these sources.  

The jets can at the simplest level, be modelled by two intrinsic properties – Lorentz factor (${\gamma}$) and angle to the line of sight (${\theta}$). Homan (2011) noted that by studying the apparent speed distribution of a large, flux-density limited sample of AGN jets, we probe the underlying Lorentz factor distribution of the parent population, assuming that the apparent speeds of moving jet components are good tracers of the underlying flow velocity of the beam (Vermeulen \& Cohen 1994; Lister \& Marscher 1999).
\section{Theory of Relationship}
\label{sect:Obs}
In the relativistic beam model of extragalactic radio sources (EGRS), the cores of these sources are believed to be responsible for the ejection of the observed VLBI jet components, and thus observed properties are expected to correlate. As these components are ejected, they move away from the core. Assuming ballistic motion, the observed radial distance $(D)$ from the core should be fore-shortened due to combined relativistic motion and orientation effect, thus we can write (see Ubachukwu 1998)
\begin{equation}
D=D_0{\mathrm{sin}}{\theta},
\end{equation}
where $D_0$ is the intrinsic length of the jet component from the core in its rest frame, and ${\theta}$ the viewing angle. The observed apparent speed ${\beta}_{\mathrm{a}}$ with which these components move out from the core from orientation based argument, can be defined as 
\begin{equation}
{\beta}_{\mathrm{a}}={\gamma}{\beta}{\delta}{\mathrm{sin}}{\theta},
\end{equation}
where ${\delta}$ is the Doppler factor defined by  ${\delta}= {\gamma}^{-1}{(1- {\beta} {\mathrm{cos}}{\theta})}^{-1}$, $ {\gamma}$ is the Lorentz factor, which is related to  ${\beta}$  the bulk speed defined in units of $c$ (the speed of light) by ${\gamma} =(\sqrt{1-{{\beta}}^2})^{-1}$. The maximum apparent speed $({\beta}_{\mathrm{a}})_{\mathrm{max}}$ for a given Lorentz factor ${\gamma}$ occurs at a critical angle  ${\theta}_c$ given by ${\beta}={\mathrm{cos}}{\theta}_c$  which implies that ${\mathrm{sin}}{\theta}_c={\gamma}^{-1}$. The Doppler factor for a given Lorentz factor is maximum for  ${\beta}= {\mathrm{cos}}{\theta}$ and is given by ${\delta}={\gamma}$.  Equation (2) can thus be written approximately as,
\begin{equation}
({\beta}_{\mathrm{a}})_{\mathrm{max}}{\approx}{\beta}{\gamma}.
\end{equation}
Thus, the time $(t_{\mathrm{D}})$ taken for the component to move away from the core to an observed maximum radial distance $D_{\mathrm{max}}$ is given by 
\begin{equation}
t_{\mathrm{D}}=\frac{D_{\mathrm{max}}}{{\beta}{\gamma}}.
\end{equation}
Kovalev et al. (2005) noted that most new jet features typically increase in size and/or decrease in flux density as a result of adiabatic expansion and/or synchrotron losses. Thus, as these components move away from the stationary core, they are expected to expand sideways. Assuming free adiabatic expansion, we follow (e.g. De Young 2002; K\"{o}rding \& Falcke 2004) to define the expansion speed  (which is expected to equal the sound speed) as
\begin{equation}
{\beta}_{\mathrm{s}}= \frac{1}{{\sqrt{\frac{1}{{\Gamma -1}}+{\frac{m_{\mathrm{p}}n_{\mathrm{tot}}c^2}{{\Gamma}P}}}}},
\end{equation}
where $m_{\mathrm{p}}$ is the proton mass, $n_{\mathrm{tot}}$ total number density of particle,  ${\Gamma}$ adiabatic index, $P$ is the thermodynamic equilibrium pressure. The adiabatic index is usually taken to be ~4/3 or ~5/3 for relativistic and non relativistic expansion respectively. K\"{o}rding \& Falcke (2004) showed that for photon emitting plasma (which is what we expect from a radio component) the maximal expansion speed will be
\begin{equation}
{\beta}_{\mathrm{s,max}}= {\sqrt{{\Gamma -1}}}
\end{equation}
Thus, the time $(t_{\mathrm{R}})$ taken for the jet component to reach an observed maximum size $R_{\mathrm{max}}$ is given by
\begin{equation}
t_{\mathrm{R}}=\frac{R_{\mathrm{max}}}{\sqrt{{\Gamma -1}}}.
\end{equation}
 For a free ballistic adiabatic expanding gas in isotropic environment, with constant expansion speed and linear speed away from the core, the time to expand to the maximum size $t_{\mathrm{R}}$, should correlate with the time to reach the maximum radial distance $t_{\mathrm{D}}$. Thus, from equations ~(4) and ~(7) we have
\begin{equation}
\frac{R_{\mathrm{max}}}{\sqrt{{\Gamma -1}}}= \frac{D_{\mathrm{max}}}{\sqrt{{\gamma}^2 -1}}.
\end{equation}
By implication, a linear regression fit to equation ~(8) in the form 
\begin{equation}
{\log}D_{\mathrm{max}}={\log}R_{\mathrm{max}} + {\log}\sqrt{\left({\frac{{\gamma}^2 -1}{{{\Gamma -1}}}}\right)}.
\end{equation}
will enable us estimate the average Lorentz factor and thus place some useful constraint on the angle to the line of sight of the observer.  Also, the ${\log}D_{\mathrm{max}} - {\log}R_{\mathrm{max}}$ plot is expected to yield a slope of unity, which can be tested using a well-defined source sample. We chose a logarithmic form of the relationship due to the wide spread in $R_{\mathrm{max}}$ and $ D_{\mathrm{max}}$.
\section{Data Analysis/Result}
Our analyses were based on the Caltech – Jodrell Bank Flat Spectrum Sources (CJF) defined by Taylor et al. (1996) which is a complete flux-limited VLBI sample of 293 flat spectrum radio sources drawn from the 6 cm and 20 cm Green Bank Surveys. Of the 293 sources in the original sample, we obtained information on 799 jet components for 237 sources from Britzen et al. (2007, 2008), with at least one observed component apparent velocity for each source, and other parameters  like proper motion,  distance of each component with respect to a stationary core and sizes (major axis) of each component. This sample is large, observational strategy and data reduction is homogeneous. Thus, given its size, completeness, and range of observed source properties, it will allow a meaningful statistical investigation of apparent motions and many correlations of other parameters. 

In our analyses, we assumed that a source contains an ideal beam $-$ one that is straight and narrow, with the pattern speed ${\beta}_{\mathrm{p}}$ equal to the bulk speed  ${\beta}_{\mathrm{b}}$.  Many sources however, are seen to have more than one moving components, with different values of apparent transverse speed ${\beta}_{\mathrm{a}}$. We selected the component with the fastest speed for sources with more than one component, believing that it is the component most likely to have the same speed as the beam (see Cohen et al. 2007) and would have reached a maximum distance for a given time. More also, Kellermann et al. (2004) suggested that there is a characteristic speed for each jet which is probably reflected by the fastest speed observed for the jet component. The final sample consists of ~177 quasars, ~41 galaxies and ~19 BL Lac objects. However, Britzen et al. (2008) noted that selecting the components based on the brightest components is a good representative of the sample and adequate for population studies.  Thus, in this paper, we will compare our results obtained from selection based on fastest components, with that based on brightest components. We point out that sources with only one observed component, is represented in both selections by that component.

Assuming the radius of the semi-major axis  of each source as representing the size, and transforming the angular distance of each component from the stationary core to a linear distance, for the selection based on fastest/brightest components, we estimated the Lorentz factor of each source from equation ~ (8) assuming $4/3 \le {\Gamma} \le 5/3$. The histogram plots of the estimated Lorentz factor for each source in each class of object are shown in figures $1 - 6$. Generally, the distributions are similar for ${\Gamma}=4/3$ and ${\Gamma}=5/3$ for each class of object, though shifted to higher values for  ${\Gamma}=5/3 $ and are generally positively skewed. The median value of Lorentz factor estimated for each class of object for $4/3 \le {\Gamma} \le 5/3$, for selection based on fastest components are  $\sim 9\pm2 -11\pm2$; $\sim 6\pm1 - 10\pm2$; and $\sim 4\pm1 - 6\pm1$ respectively for quasars, galaxies and BL Lac objects. Similarly, for the selection based on brightest components we have $\sim 6\pm1 - 10\pm2$; $\sim 5\pm1 - 8\pm1$ and $\sim 5\pm1 - 8\pm1$ for quasars, galaxies and BL Lac objects respectively for $4/3 \le {\Gamma} \le 5/3$. The quoted errors were estimated from the observed error associated with the parameters obtained from  Britzen et al. (2007, 2008).

\begin{figure}[h]
  \begin{minipage}[t]{0.495\linewidth}
  \centering
   \includegraphics[width=80mm,height=60mm]{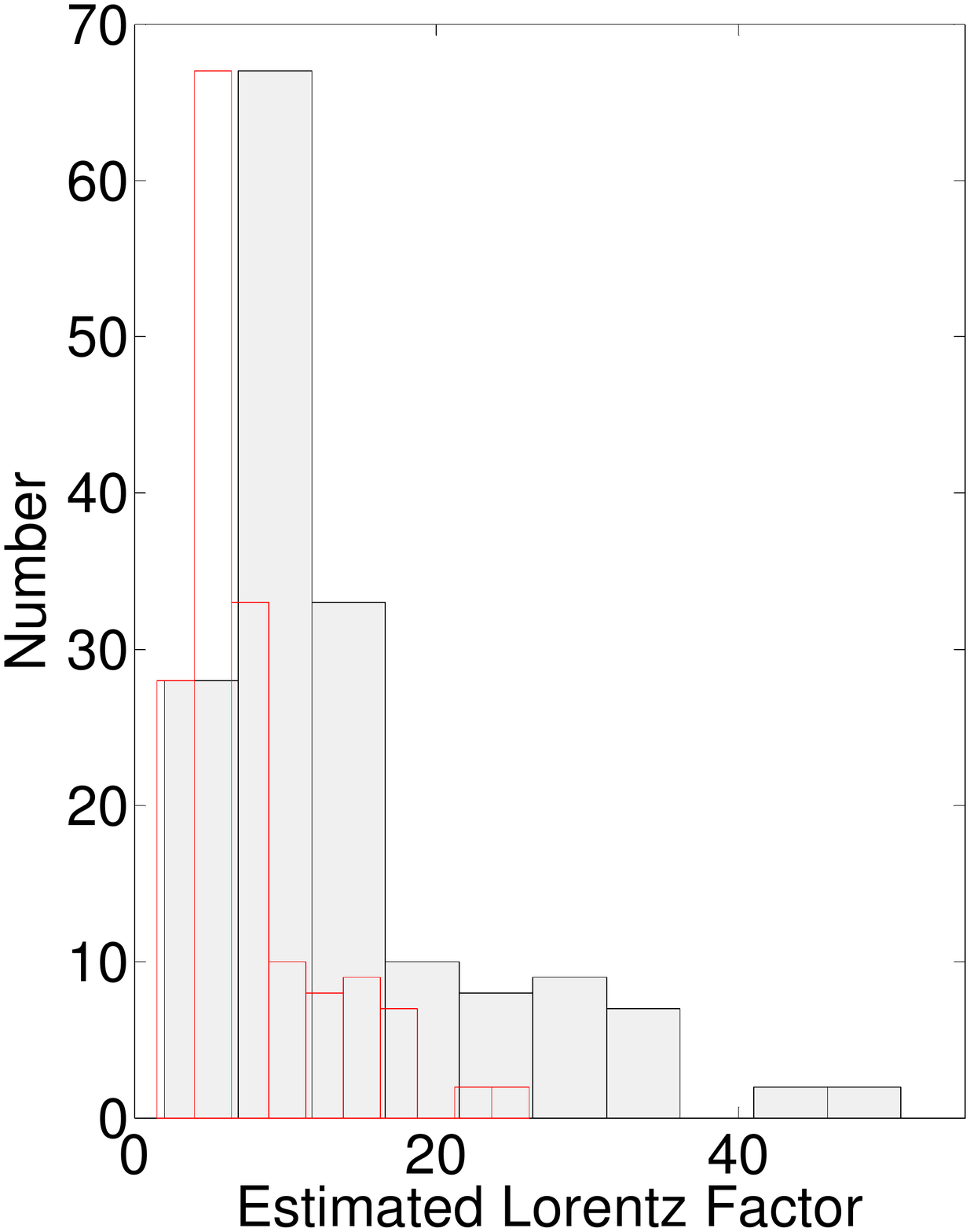}
   \caption{{\small  Histogram Plot of Estimated Lorentz Factor for the selection based on fastest components for Quasars (Grey Plot for ${\Gamma}= 5/3$; Red line for ${\Gamma}= 4/3$)} }
  \end{minipage}%
  \begin{minipage}[t]{0.495\textwidth}
  \centering
   \includegraphics[width=80mm,height=60mm]{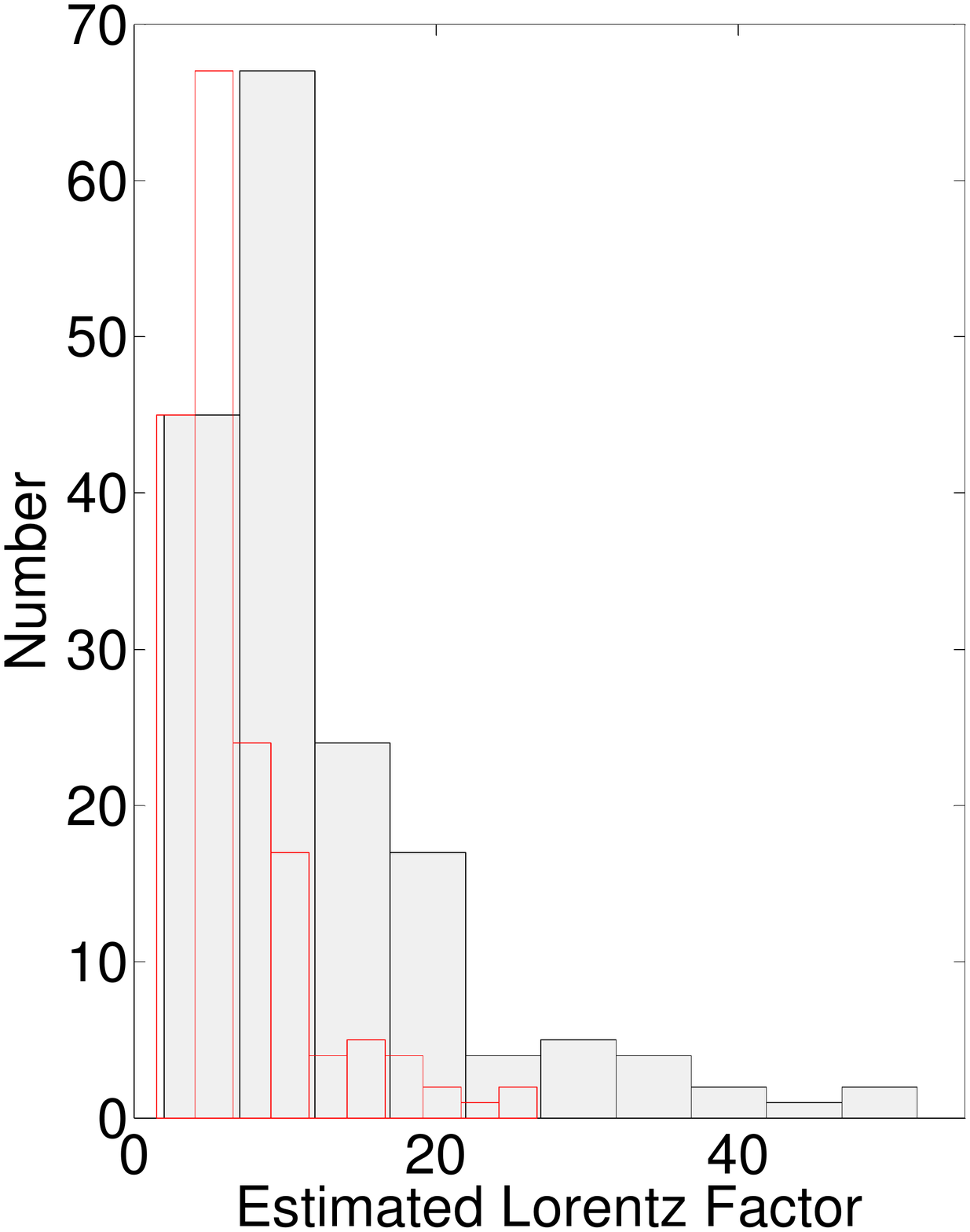}
  \caption{{\small Histogram Plot of Estimated Lorentz Factor  for the selection based on brightest components for Quasars (Grey Plot for ${\Gamma}= 5/3$; Red line for ${\Gamma}= 4/3$)}}
  \end{minipage}%
  \label{Fig:fig12}
\end{figure}

\begin{figure}[h]
  \begin{minipage}[t]{0.495\linewidth}
  \centering
   \includegraphics[width=80mm,height=60mm]{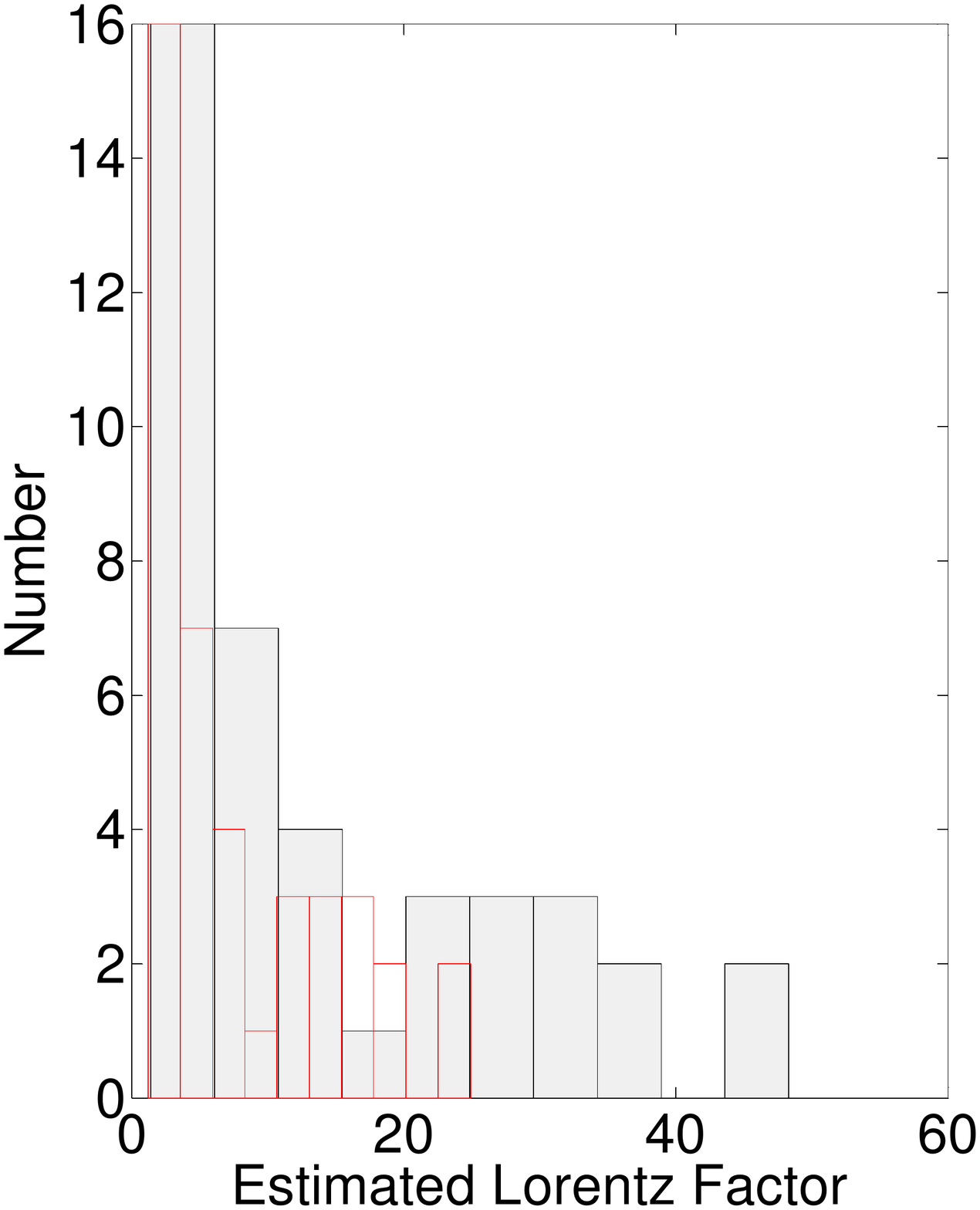}
   \caption{{\small Histogram Plot of Estimated Lorentz Factor for the selection based on fastest components for Galaxies (Grey Plot for ${\Gamma}= 5/3$; Red line for ${\Gamma}= 4/3$)} }
  \end{minipage}%
  \begin{minipage}[t]{0.495\textwidth}
  \centering
   \includegraphics[width=80mm,height=60mm]{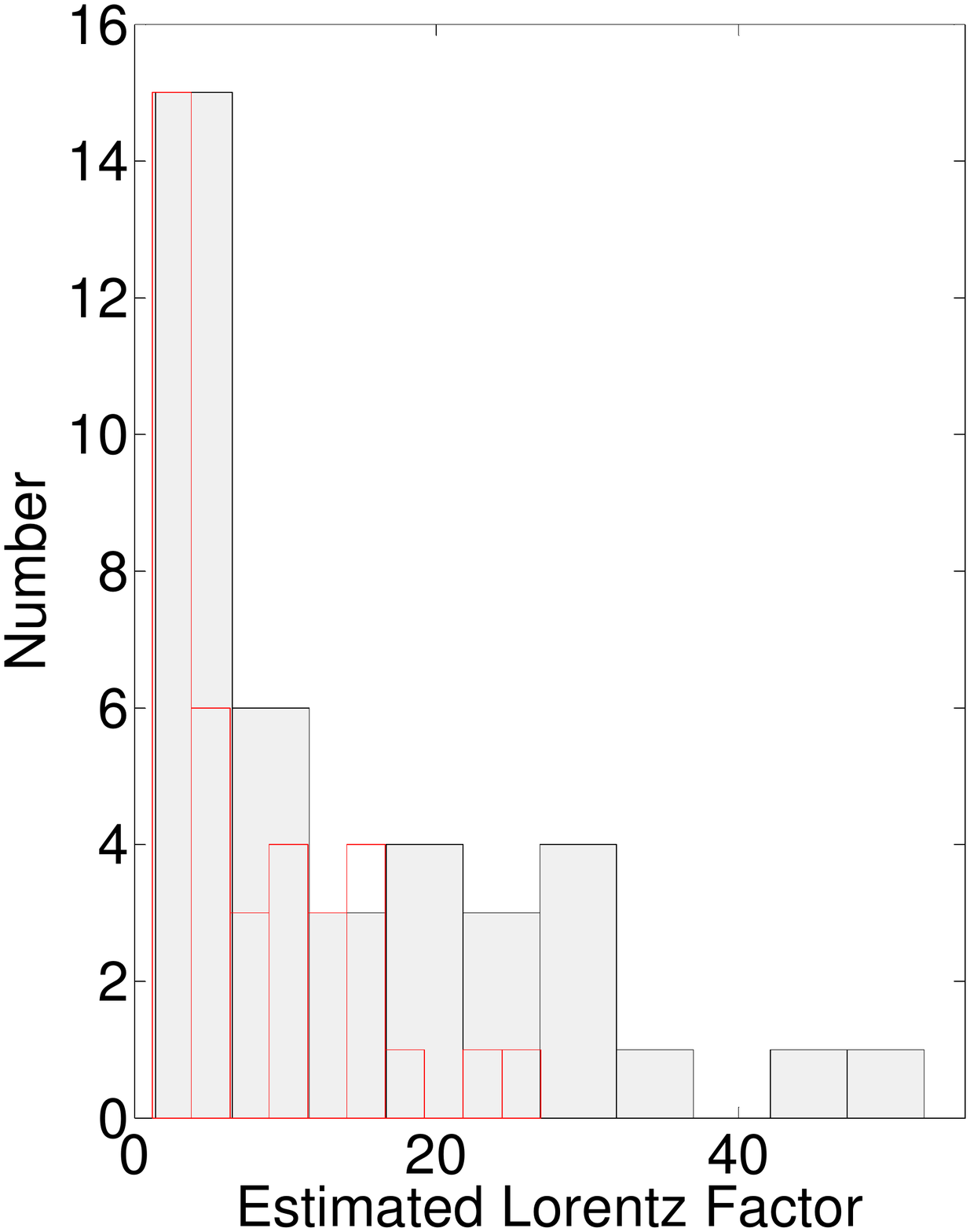}
  \caption{{\small Histogram Plot of Estimated Lorentz Factor for the selection based on brightest components for Galaxies (Grey Plot for ${\Gamma}= 5/3$; Red line for ${\Gamma}= 4/3$)}}
  \end{minipage}%
  \label{Fig:fig34}
\end{figure}

\begin{figure}[h]
  \begin{minipage}[t]{0.495\linewidth}
  \centering
   \includegraphics[width=80mm,height=60mm]{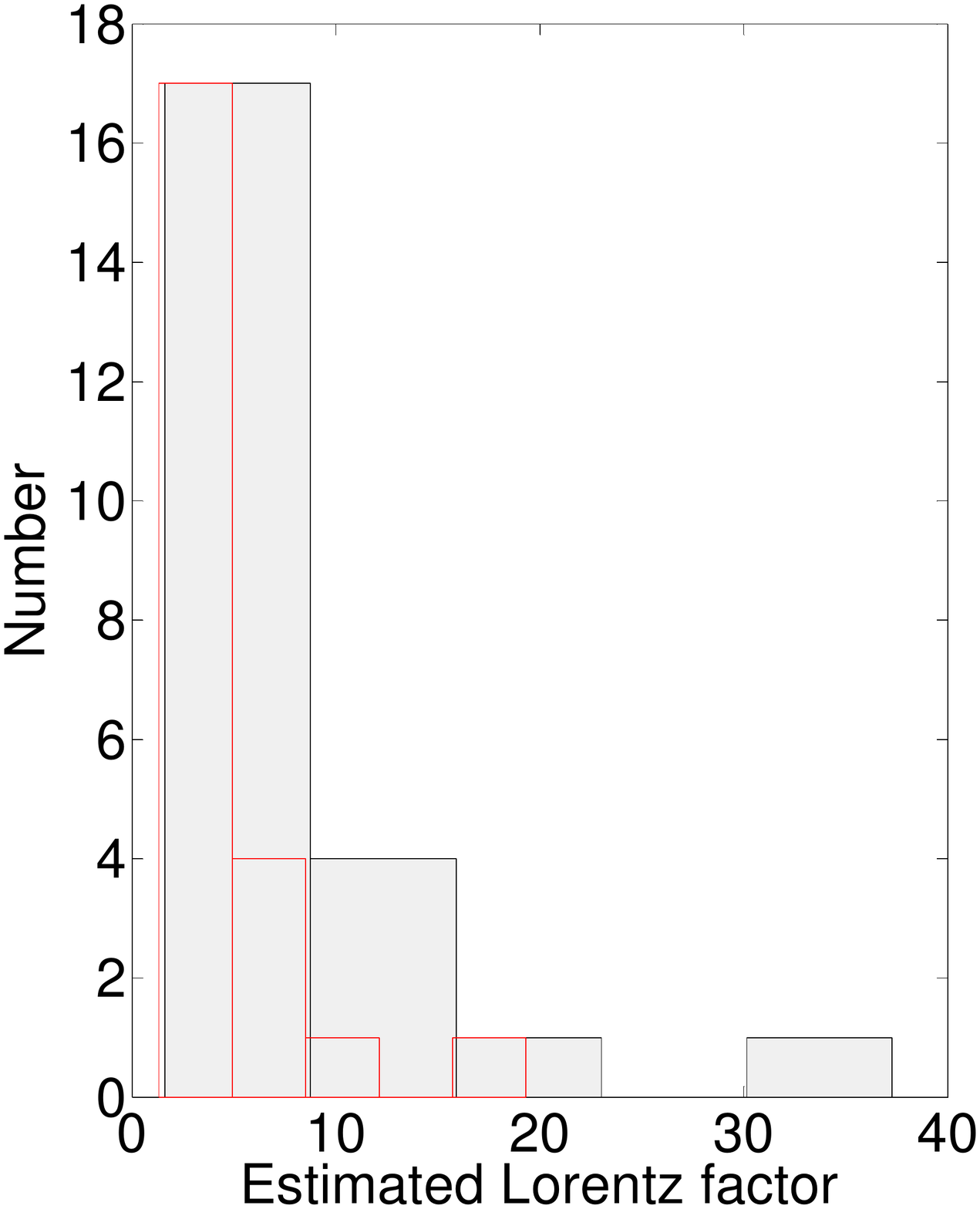}
   \caption{{\small Histogram Plot of Estimated Lorentz Factor for the selection based on fastest components for BL Lac (Grey Plot for ${\Gamma}= 5/3$; Red line for ${\Gamma}= 4/3$)} }
  \end{minipage}%
  \begin{minipage}[t]{0.495\textwidth}
  \centering
   \includegraphics[width=80mm,height=60mm]{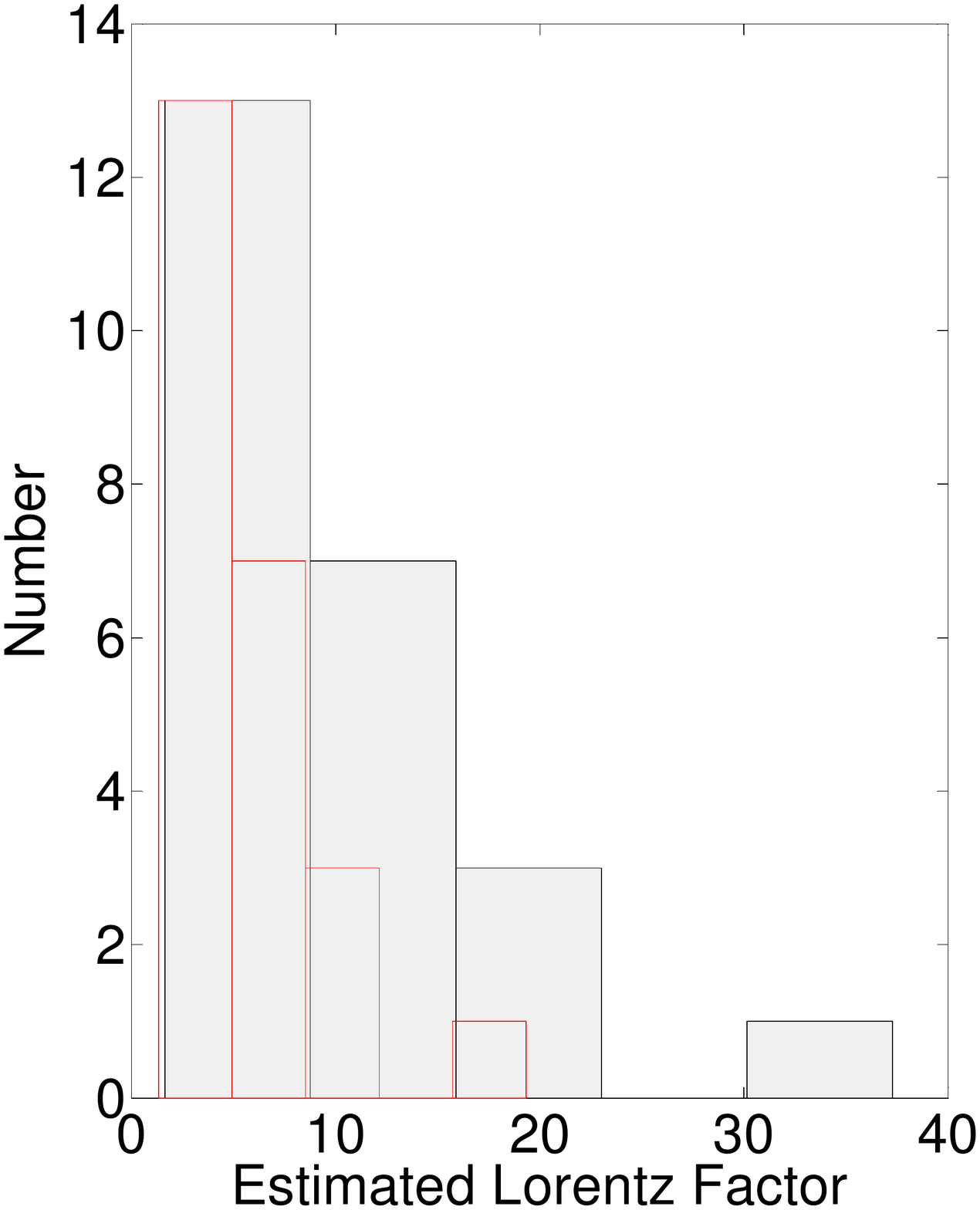}
  \caption{{\small Histogram Plot of Estimated Lorentz Factor assuming for the selection based on brightest components for BL Lac (Grey Plot for ${\Gamma}= 5/3$; Red line for ${\Gamma}= 4/3$)}}
  \end{minipage}%
  \label{Fig:fig56}
\end{figure}

We also fitted the observed $ D-R$ data to equation ~ (9). The result gives $: \log{D} = (0.7 {\pm}0.3){\log}R + 1.2{\pm}0.1$ for quasars; $ \log{D} = (0.3 {\pm}0.5){\log}R + 1.1{\pm}0.1$ for galaxies; and $\log{D} = (0.6 {\pm}0.4){\log}R + 0.9{\pm}0.1$ for BL Lac objects. This corresponds to a distribution in Lorentz factor of: ${\gamma} \sim 7-17$ for quasars; ${\gamma} \sim 6-13$ for galaxies; and ${\gamma} \sim 4-9$ for BL Lac objects  for $ \frac{4}{3}{\le}{\Gamma}{\le}\frac{5}{3}$, with correlation coefficient between $ D-R$ as  $r \sim 0.6, 0.4, $ and $0.8$ respectively; a fairly strong correlation, especially for quasars and BL Lac objects. The spread in the Lorentz factor results also imply a distribution in the critical angle to the line of sight ${\theta}_c$ of $ \sim 3^0 - 8^0,  4^0  - 10^0, $ and $ 6^0 -16^0 $ for quasars, radio galaxies and BL Lacs respectively.

Similar analyses using selection based on brightest components yield: ${\log}{D} = (0.7 {\pm}0.3){\log}R + 1.1{\pm}0.1$ for quasars; ${\log}{D} = (0.9 {\pm}0.4){\log}R + 1.0{\pm}0.1$ for galaxies; and ${\log}{D} = (0.1 {\pm}0.4){\log}R + 0.5{\pm}0.2$ for BL Lac objects, with correlation coefficients $ r \sim 0.6, 0.5 $ and $ 0.3$ respectively. The corresponding Lorentz factor for different classes of sources are: ${\gamma} \sim 5-12$ for quasars; ${\gamma} \sim 5-9$ for galaxies; and ${\gamma} \sim 2-5$ for BL Lac objects  for $ \frac{4}{3}{\le}{\Gamma}{\le}\frac{5}{3}$, corresponding to a distribution in  ${\theta}_c $ of $\sim 5^0- 10^0; \sim 6^0 - 12^0;$ and $\sim 12^0 - 36^0$ for quasars, radio galaxies and BL Lac objects respectively. The plot of ${\log}D_{\mathrm{max}}$ against ${\log}R_{\mathrm{max}}$  for the fastest components is shown in figure ~7, while the plot of ${\log}D_{\mathrm{max}}$ against ${\log}R_{\mathrm{max}}$  for the brightest components is shown in figure ~8.  In general, the Lorentz factor obtained using the fastest components is greater than that obtained using the brightest components. We emphasis that values estimated using the selection based on brightest components may represent the lower limits to the Lorentz factors for each class of object, and consequently the upper limit to the viewing angle.

\begin{figure}[h] 
\includegraphics[width=160mm,height=120mm]{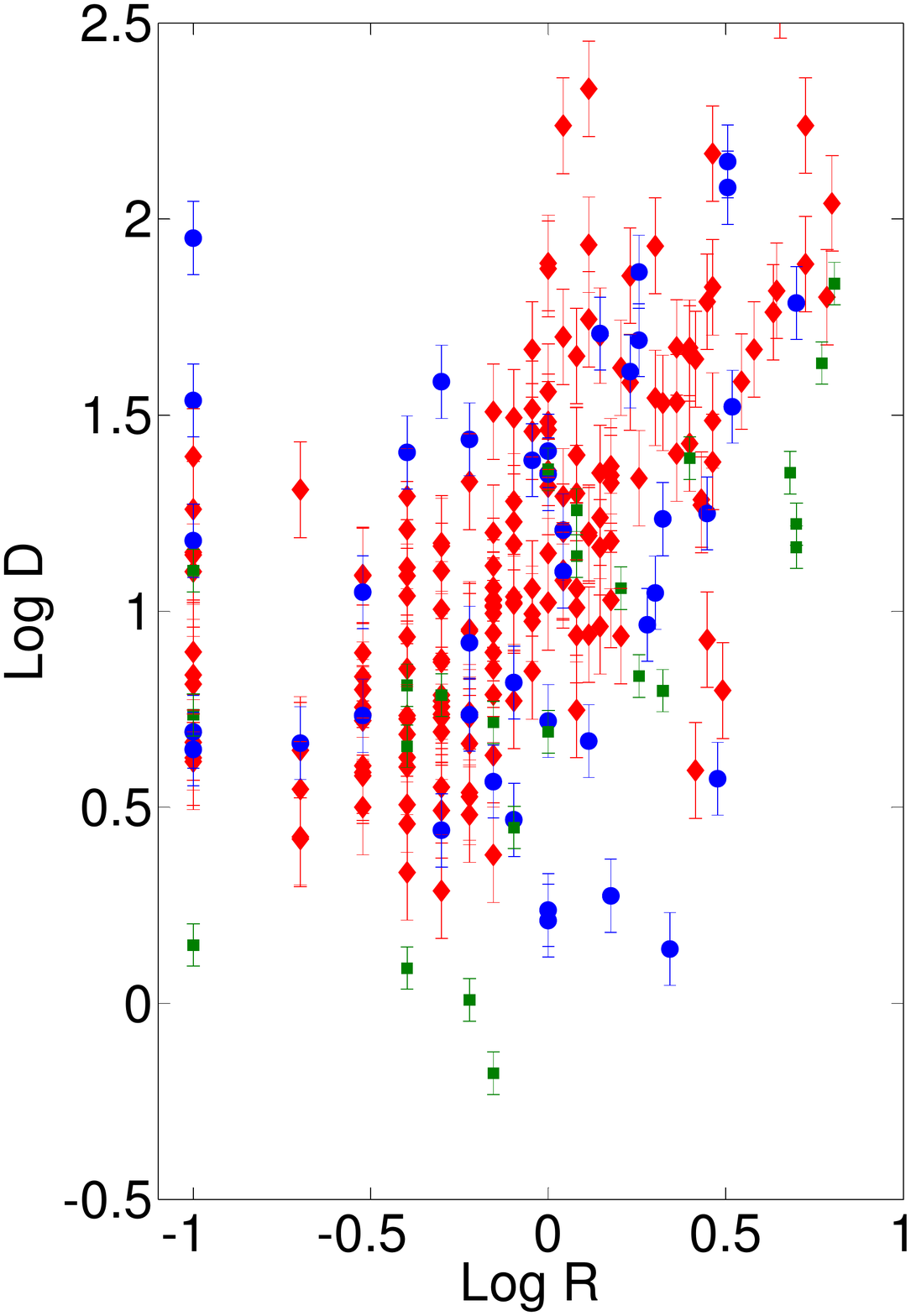}
\caption{Plot of ${\log}{\mathrm{D}}_{\mathrm{max}}$ against ${\log}{\mathrm{R}}_{\mathrm{max}}$  for the fastest components with error bar in standard deviation (Legend:Quasars - red diamond; Galaxies - blue circle; BL Lacs - green square)}
\label{Fig:fig3}
\end{figure}
\begin{figure}[h] 
\includegraphics[width=160mm,height=120mm]{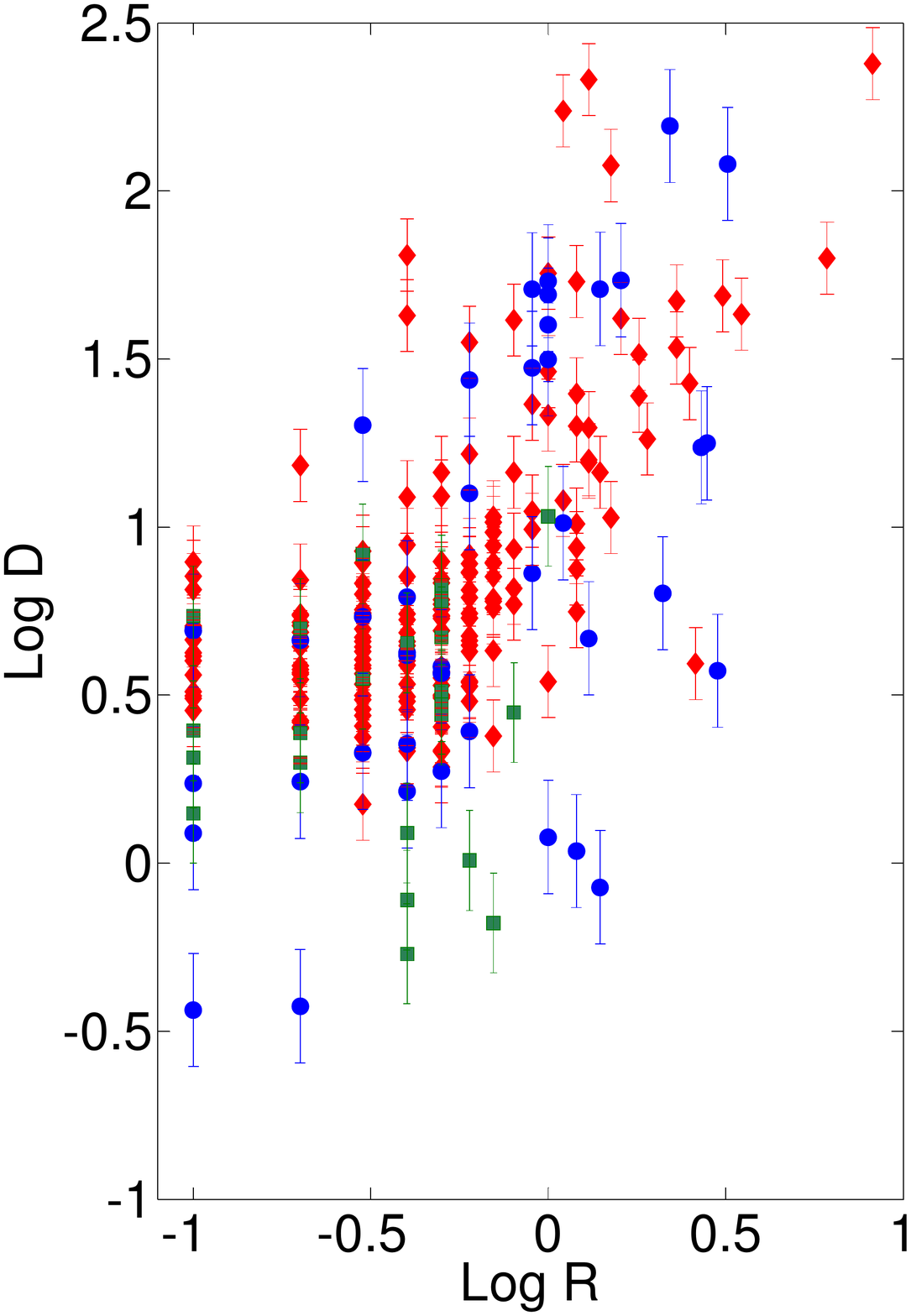}
\caption{Plot of ${\log}{\mathrm{D}}_{\mathrm{max}}$ against ${\log}{\mathrm{R}}_{\mathrm{max}}$  for the brightest components with error bar in standard deviation (Legend:Quasars - red diamond; Galaxies - blue circle; BL Lacs - green square)}
\label{Fig:fig4}
\end{figure}

\section{Discussion and Conclusion}
We have used the relationship between the observed component radial distance $D$ and component size $R$ to estimate the Lorentz factor $({\gamma})$ and the corresponding angle to the line of sight ${\theta}$ for a selection based on fastest/brightest components. Our results show a stronger $D-R$ correlation for selection based on fastest components than for that based on brightest components for quasars, radio galaxies and BL Lac sub-samples. Our results also indicate lack of apparent correlation between $D$ and $R$ in the BL Lac data for selection based on the brightest component. This suggests that selection based on the brightest components may not be appropriate for the determination of the Lorentz factor or for characterizing superluminal motion for this class of object. In general, the result on the analysis of CJF sample obtained here, showed that galaxies have lower intrinsic Lorentz factor (lower bulk speed) than quasars. There are few BL Lac in the sample, but the result showed that BL Lac have the lowest range of Lorentz factor in agreement with (e.g. Gabuzda et al. 1994; Morganti et al. 1995; Kellermann et al. 2007). This is an indication that BL Lac objects are low power sources (Bicknell 1994). We  note that equation ~(9) yields a slope of unity for the plot of the $R-D$ data. The results from our regression analyses approximate the theoretical value, especially for quasars, and in general for selection based on fastest components. This is an indication that the underlying assumption is plausible. These results  do show that selections based on fastest component are a good indicator of the bulk speed. 

We note that  our estimated values for angle to the line of sight especially for sample selected based on brightest components (especially for BL Lacs) are high (see  Hovatta et al. 2009; Yang et al. 2012), however, Gopal-Krishna et al. (2006) and Wiita et al. (2008) pointed out that allowing for the conical shape of ultra relativistic blazar jets with opening angles of a few degrees on parsec-scales, they showed that the actual viewing angles of these conical jets from the line-of-sight can be substantially  larger than the values usually inferred by combining their flux-variability and proper motion measurements (Jorstad et al. 2005). This jet geometry also implies that de-projected jet opening angles will typically be significantly underestimated from VLBI measurements. Also, our estimated Lorentz factor in general are lower than that shown in Table ~3 of Hovatta et al. (2009), except for galaxies, though their sample was obtained at ~22 and ~37 GHz as against the CJF sample observed at ~5 GHz. Kellermann et al. (2004) noted that there is a systematic decrease in apparent velocity with increasing wavelength which indicates that observation at different wavelengths may be sampling different part of the jet structure. Moreover, observations at lower frequency have lower angular resolution and thus more sensitive to the lower surface brightness structure located downstream.

Further analysis of our result, showed that for the sample selected based on brightest components, the apparent speed, component size, and component distance from the assumed stationary core are on the average less than that of sample selected based on fastest components, but with brighter apparent luminosity (see Table 1). They may be expected to have smaller and apparently brighter components, but if their motion is unimpeded, they should also on the average, have higher apparent speed. This is an indication that their apparent luminosity might be more of projection effect/motion through bends than being relativistically Doppler boosted. Thus, Lorentz factor calculated based on brightest components for a given sample, will most likely represent the lower limit to the bulk expansion speed, and upper limit to the viewing angle, while that estimated from the fastest components will represent the most probable values (Gopal-Krishna et al. 2006; Marshall et al. 2011). In Hovatta et al. (2009), their Figure 11 shows that ${\gamma}{\mathrm{sin}}{\theta}$ distribution for their sources peaks  around ~1, though a number of sources have  ${\gamma}{\mathrm{sin}}{\theta}$  larger or smaller, thus the assumption of ${\mathrm{sin}}{\theta}=1/{\gamma}$ seems reasonable.

\begin{table}[h]
\caption{Comparison of the Mean values of some parameters. (Between sample selected based on fastest component and that selected based on brightest components)}
\begin{tabular}{|llllll|} \hline
 &  & ${\beta}_{\mathrm{a}}$ & Component size  & Component Luminosity & Component Distance \\
 & & &$ (mas) $ & $ (Wm^{-2}Hz^{-1}) {\mathrm{x}}10^{25}$ & (pc) \\\hline
Quasars & fastest & 9.43 & 1.14 & 4.16 & 24.59\\
& brightest & 5.40 & 0.69 & 6.87 & 13.65\\\hline
Galaxies & fastest & 4.43 & 1.30 & 1.45 & 25.26\\
& brightest & 2.63 & 0.95 & 2.28 & 20.00\\\hline
BL Lacs & fastest & 4.52 & 1.35 & 2.40 & 10.09\\
& brightest & 3.45 & 0.44 & 3.33 & 3.73\\\hline
\end{tabular}
\end{table}

Cohen et al. (1988) used a sample of compact radio quasars with measured redshift ($z$) and proper motion ${\mu}$ to obtained a result which indicated that assumption of relativistic beaming model gives a reasonable fit to their data with an upper bound of bulk Lorentz factor lying in the range $9\le{\gamma}\le18$. Pelletier \& Roland (1989) believed superluminal motion in powerful radio sources can be explained by a two – fluid model. In their modelling they estimated that relativistic electrons responsible for the synchrotron emission of VLBI jets are moving with Lorentz factor $3\le{\gamma}\le10$. In the analysis of flux variations in BL Lac objects, Mutel (1992) indicated that the Lorentz factor lies in the range $2\le{\gamma}\le4$. For a sample of 100 sources with published VLBI measurements of the core angular size, Ghisellini et al. (1993) assuming Synchrotron Self – Compton (SSC) formalism estimated the Doppler factor of these sources independent of superluminal motion. Using the estimated Doppler factor and observed superluminal speed ${\beta}_{\mathrm{a}}$ they constrained ${\gamma}=10$. In VLBI observations of ~12 Flat Spectrum Radio Quasars (FSRQ) in ~2 Jy sample, Vermeulen \& Cohen (1994) obtained a  distribution of  ${\gamma}$ for the fastest superluminal components in FSRQ in the range of $5 - 35$. Urry \& Padovani (1995) showed that for their derived luminosity function of FSRQ to match the observed FSRQ luminosity function requires a distribution in Lorentz factor in the range of $5 \leq{\gamma}\leq40$ with a mean value of $\langle{\gamma}\rangle=11$. While for BL Lac objects, a range of $2\le{\gamma}\le20$ and an average of $\langle{\gamma}\rangle=3$ was needed for their luminosity function to match the observed value.  They also obtained a critical angle value for maximum beaming of ${\theta}_c=14^0$ for FSRQ and ${\theta}_c=12^0$ or ${\theta}_c=19^0$ for BL Lac depending on the assumed model. These results generally, agree with our result, especially the result obtained with sample selected based on fastest components. The seemingly contradictory result for the BL Lac sample based on brightest component may be interpreted as motion through bends with increase in brightness due to projection effects. The data are however too poor for any definite conclusion.

Miller-Jones et al. (2004) pointed out that for tangled magnetic field $ B{\propto} R^{-1}$, thus with both synchrotron loses and adiabatic loses due to expansion, as the component moves out, size increases, thus we expect component luminosity to anti-correlate with both component size and component distance, except modified by projection/environmental effects. Homan et al. (2002) found that flux density of jet components fades from the core as $D^{-1.3}$. These are in support of our argument that time for component expansion should correlate with period the components have move a radial distance away from the core. The correlation analysis between component size $R$  and component luminosity $L$ shows $R - L^{-x}$, for the selection based on fastest components with correlation coefficient results are   $r = -0.3,-0.2$ and $-0.3$ for quasars, galaxies and BL Lac respectively. These results indicates that selections based on fastest component are good indicator of the bulk speed, though without any high level of significance. 

We have shown that expected increase in size as jet components move away from the core can be used to characterize the kinematics of these sources. This is true for free expanding jets beyond the collimation point. Jets observed at high frequency may be sampling the upstream part (see Kellermann et al. 2004), thus a re-analysis of VLBI sources observed at different frequency range may be used to place a limit on the point of collimation.

\begin{acknowledgements}
This work received funding from TEF and we also acknowledge the contributions of the anonymous referee. 
\end{acknowledgements}


\begin{thebibliography}{}
\bibitem[Bicknell (2007)]{bic94} Bicknell G. V., 1994, \apj, 422, 542
\bibitem[Britzen et al. (2007)]{bri07} Britzen, S., Vermeulen, R.C., Taylor, G.B., Campbell, R.M., Pearson, T.J., Readhead, A.C.S., Xu, W., Browne, I.W.A., Henstock, D.R., \& Wilkinson, P. 2007, \aap, 472, 763
\bibitem[Britzen et al. (2008)] {bri08} Britzen, S., Vermeulen, R.C., Campbell, R.M., Taylor, G.B., Pearson, T.J., Readhead, A.C.S., Xu, W., Browne, I.W.A., Henstock, D.R., \& Wilkinson, P., 2008,\aap, 484, 119
\bibitem[Cohen et al. (1988)]{coh88} Cohen, M.H., Barthel, P.D., Pearson, T.J., \& Zensus, J.A., 1988, \apj, 329, 1
\bibitem[Cohen et al. (2007)]{coh07} Cohen, M. H., Lister, M. L., Homan, D. C., Kadler, M., Kellermann, K. I., Kovalev, Y. Y., \& Vermeulen, R. C. 2007, \apj, 658, 232
\bibitem[De Young (2002)]{you02} De Young, D.S., 2002, The Physics of Extragalactic Radio Sources, University of Chicago Press Limited. Chicago. pp549
\bibitem[Gabuzda et al. (1994)]{gab94} Gabuzda, D.C., Mullan, C.M., Cawthorne, T.V., Wardle, J.F.C., \& Roberts, D.H., 1994, \apj, 435, 144
\bibitem[Ghesellini et al. (2007)]{ghe93} Ghisellini, G., Padovani, P., Cellotti, A., \& Maraschi, L., 1993, \apj,  407, 65
\bibitem[Gopal-Krishna et al. (2006)]{gop06} Gopal-Krishna, Wiita, P.J., \& Dhurde, S., 2006, \mnras, 369, 1287
\bibitem[Homan (2011)]{hom04} Homan, D.C., 2011, pre-print (astro-ph 1110.4852v1) 1
\bibitem[Homan et al. (2002)]{hom02} Homan, D.C., Ojha, R., Wardle, J.F.C., Roberts, D.H., Aller, M.F., Aller, H.D., \& Hughes, P.A., 2002, \apj, 568, 99
\bibitem[Hovatta et al. (2009)]{hov09} Hovatta, T., Valtaoja, E., Tornikoski, M., \& L\"{a}hteem\"{a}ki, A., 2009, \aap, 494, 527
\bibitem[Jorstad et al. (2005)]{jor05}Jorstad, S.G.,  Marscher, A.P., Lister, M.L.,  Stirling, A.M., Cawthorne, T.V.,  Gear, W.K., Go\a'mez, J. L.,  Stevens, J.A., Smith, P.S., Forster, J.R., \& Robson I. E.,  2005, \aj, 130, 1418
\bibitem[Kellermann et al. (2007)]{ker07} Kellermann, K. I., Kovalev, Y. Y., Lister, M. L., Homan, D. C., Kadler, M., Cohen, M. C., Ros, E., Zensus, J.A., Vermeulen, R.C., Aller, M.F., \& Aller H.D., 2007, pre-print (arXiv:0708.3219v1)
\bibitem[Kellermann et al. (2004)]{ker04} Kellermann, K. I., Lister, M. L., Homan, D. C., Vermeulen, R.C., Cohen, M. C., Ros, E., Kadler, M., Zensus, J.A., \& Kovalev, Y. Y., 2004, \apj, 609, 539
\bibitem[Kovalev et al. (2005)]{ker05} Kovalev, Y. Y., Kellermann, K. I., Lister, M. L., Homan, D. C., Vermeulen, R.C., Cohen, M. C., Ros, E., Kadler, M., Lobanov, A.P., Zensus, J.A., Kardashev, N.S., Gurvits, L.I., Aller, M.F., \& Aller H.D., 2005, \aj, 130, 2473
\bibitem[Kording \& Falcke (2004)]{kor04} K\"{o}rding, E., \& Falcke, H., 2004 in the role of VLBI in Astrophysics Astronomy and Geodesy eds. F. Montavani and A. Kus,  Kluwer Academic Publishers, Netherlands, 107-127
\bibitem[Lister \& Marscher (1999)]{lis99} Lister, M.L., \& Marscher, A.P., 1999,  Astroparticle Physics 11, 65 
\bibitem[Marshall et al. (2011)]{mar04}Marshall, H. L., Gelbord, J. M., Schwartz, D. A.,  Murphy, D. W., Lovell, J. E. J.,  Worall, D. M., Birkinshaw, M., Perlman, E., Godfrey, S. L., \& Jauncey, D. L., 2011, \apjs, 193, 15
\bibitem[Miller-Jones et al. (2004)]{mil04} Miller-Jones, J.C.A., Blundell, M.K., \& Duffy, P., 2004, \apj, 603, L21
\bibitem[Morganti et al. (1995)]{mor95} Morganti, R., Oosterloo, T.A., Fosbury, R.A., \& Tadhunter, C.N., 1995. \mnras, 274, 393M
\bibitem[Mutel (1992)]{mut92} Mutel , R. l., 1992, Extragalactic Radio Sources. From Beams to Jets. Proceedings of the 7th. I.A.P.  Eds, J. Roland, H. Sol, G. Pelletier; Publisher, Cambridge University Press, 1992. ISBN  0-521-41602-7. LC QB463 .I56 1991, p. 130 – 133
\bibitem[Pelletier \& Roland (1989)]{pel89} Pelletier, G., \& Roland, J., 1989, \aap, 224, 24
\bibitem[Taylor et al. (1996)] {tay96} Taylor, G.B., Vermeulen, R.C., Readhead, A.C.S., Pearson, T.J., Henstock, D.R., \& Wilkinson, P.N., 1996, \apj, 107, 37
\bibitem[Ubachukwu (1998)]{uba98}Ubachukwu, A.A., 1998, Ap\&SS, 257, 23
\bibitem[Urry \& Padovani (1995)]{urr95} Urry, C.M., \& Padovani, P., 1995, \pasp, 715 
\bibitem[Vermeulen \& Cohen (1994)]{ver94} Vermeulen, R.C., \& Cohen, M.H., 1994, \apj, 430
\bibitem[Wiita et al (2008)]{wii08} Wiita,P.J.,  Gopal-Krishna,  Dhurde, S., \& Sircar, P. 2008 , Extragalactic Jets: Theory and Observation from Radio to Gamma Ray ASP Conference Series, Vol. 386, (Eds T. A. Rector and D. S. De Young).
\bibitem[Yang et al (2012)]{yan12} Yang, J., Fan, J., Yuan, Y.,  2012, Science China Phys, Mech\& Astro, 1510

 \end{thebibliography}
\end{document}